\documentclass[letterpaper]{article} 
\usepackage[draft]{aaai2027}  
\usepackage[hyphens]{url}  
\usepackage{comment}
\usepackage{graphicx} 
\usepackage{natbib}  
\usepackage{caption} 
\usepackage{algorithm}
\usepackage{algorithmic}

\usepackage{newfloat}
\usepackage{listings}
\DeclareCaptionStyle{ruled}{labelfont=normalfont,labelsep=colon,strut=off} 
\floatstyle{ruled}
\newfloat{listing}{tb}{lst}{}
\floatname{listing}{Listing}

\usepackage{booktabs}

\usepackage{multirow} 
\usepackage{amsfonts} 
\usepackage{colortbl} 

\title{Restorable Multilingual Voice Anonymization via Audio Watermarking}
\author{
    Haoyuan Zheng\textsuperscript{\rm 1},
    Guang Hua\textsuperscript{\rm 2},
    Xiaoxiao Miao\textsuperscript{\rm 1}\corresponding\\
}
\affiliations{
    \textsuperscript{\rm 1}Duke Kunshan University\\
    \textsuperscript{\rm 2}Singapore Institute of Technology\\

    \{haoyuan.zheng, xiaoxiao.miao\}@dukekunshan.edu.cn
}

\begin{document}

\maketitle

\begin{abstract}
Voice anonymization aims to conceal speaker identity while preserving linguistic and paralinguistic information. However, certain applications also require the ability to attribute anonymized recordings to their original speakers. Existing recovery approaches often rely on specially designed reversible anonymization systems, limiting their applicability to different anonymization architectures. In this work, we propose a post-hoc framework for restorable multilingual voice anonymization that enables speaker attribution without reconstructing the original waveform or modifying the underlying anonymization systems. Specifically, we convert 192-dimensional speaker embeddings into binary identity payloads through sign quantization and embed them into anonymized speech using a segment-wise audio watermarking strategy. Recovered payloads are subsequently matched against enrollment speaker templates for identity attribution. We evaluate the framework across four languages using two distinct anonymization backbones, BM1 and T24. Under clean conditions, the approach achieves approximately 99\% exact payload recovery and 0.96\% attribution EER while largely preserving the original privacy--utility performance. Experiments further show that speaker attribution remains reliable under several audio distortions despite imperfect payload recovery, although temporal rescaling severely disrupts recovery. Additional analyses reveal that watermark robustness varies across languages, carrier durations, and anonymization backbones.
\end{abstract}


\section{Introduction}

Speech conveys not only linguistic content but also paralinguistic information, such as speaker identity, emotion, age, gender, and accent, as well as acoustic cues about the recording environment. In particular, the biometric characteristics of a voice can reveal who is speaking, raising privacy concerns when speech recordings are publicly shared. Such recordings may also be misused by voice cloning systems to generate speech that impersonates the original speaker. Complementing reactive approaches such as deepfake detection, voice anonymization offers a proactive, user-centric approach to speech privacy protection. It aims to conceal speaker identity while preserving speech intelligibility, naturalness, and emotional information to allow different downstream tasks. The VoicePrivacy Challenge (VPC) has fostered research in this area by providing standardized protocols for evaluating the privacy–utility trade-off, with its 2026 edition introducing multilingual evaluation and stronger speaker verification attackers \cite{miao:hal-05561895}.

Common anonymization approaches include disentanglement-based voice conversion (VC) anonymization and cascaded automatic speech recognition–text-to-speech (ASR–TTS) anonymization. VC-based anonymization disentangles linguistic content, prosody, and speaker identity, then modifies or replaces the speaker representation while retaining the other representations to generate anonymized speech \cite{miao2022language, webber2024voice, das2024comparing}. ASR–TTS-based anonymization transcribes the input speech and synthesizes the recognized text with a different voice, with some systems additionally using emotion information to guide synthesis or select an emotion-matched reference \cite{meyer2022speaker, lee2024voice, ferreira2026emotion}.

While voice anonymization protects speaker privacy, some applications also require the ability to trace anonymized recordings back to their original speakers under authorized conditions. For example, an institution may share anonymized interview recordings for research while retaining a mechanism to establish their source when investigating misuse or resolving disputes. This motivates recoverable speech privacy, where speaker identity is concealed in the released voice, while subsequent recovery requires authorized access to auxiliary information or a designated recovery mechanism. 

Existing studies have explored recovery at two levels: voice restoration, which reconstructs the original voice or waveform \cite{zhang2025speechguard, wang2025rano}, and speaker identity recovery, which retrieves speaker representations for attribution or identity verification \cite{ren2023speaking, ge2025proactive}. However, voice restoration typically relies on specially designed reversible anonymization systems, while existing speaker identity recovery studies have primarily investigated voice conversion or identity manipulation detection. This motivates investigating a common post-hoc identity recovery mechanism that operates on already anonymized speech without modifying the underlying anonymization models. Its effectiveness across different multilingual anonymization systems, together with its impact on privacy and utility, remains insufficiently explored.

This work focuses on identity attribution, as speaker matching is often sufficient for auditing purposes without requiring reconstruction of the original voice or waveform. To this end, we explore a simple and effective watermarking approach that embeds information about the original speaker into anonymized speech. Specifically, we extract a speaker embedding from the original speech, convert it into a binary representation through sign quantization, and embed it into the anonymized output using an audio watermarking model. An authorized party, assumed to have access to the watermark decoder and reference recordings, can recover this representation for speaker matching. Unlike conventional watermarks carrying arbitrary identifiers, our payload retains speaker-discriminative biometric information that enables direct comparison with representations extracted from independent reference recordings.

We evaluate the proposed approach following the protocol of the multilingual anonymization track of VPC 2026. We embed source-speaker information post hoc into anonymized speech generated by two representative systems: the official BM1 baseline for disentanglement-based VC anonymization \cite{miao2022language}, and a local reproduction of T24, one of the top-performing submitted systems, for cascaded ASR–TTS anonymization \cite{ferreira2026emotion}. We assess watermark recovery, speaker attribution, and the impact of watermark insertion on ASV-based privacy, linguistic intelligibility, and emotional preservation. We further examine robustness against audio transformations and discuss the privacy risks of unauthorized watermark extraction.
Our study demonstrates the feasibility of recovering speaker identity through a common post-hoc watermarking framework across different multilingual anonymization systems. It further shows that identity attribution can remain reliable despite imperfect payload recovery, while revealing vulnerability to temporal rescaling.

\begin{figure*}[t]
    \centering
    \includegraphics[width=1\linewidth]{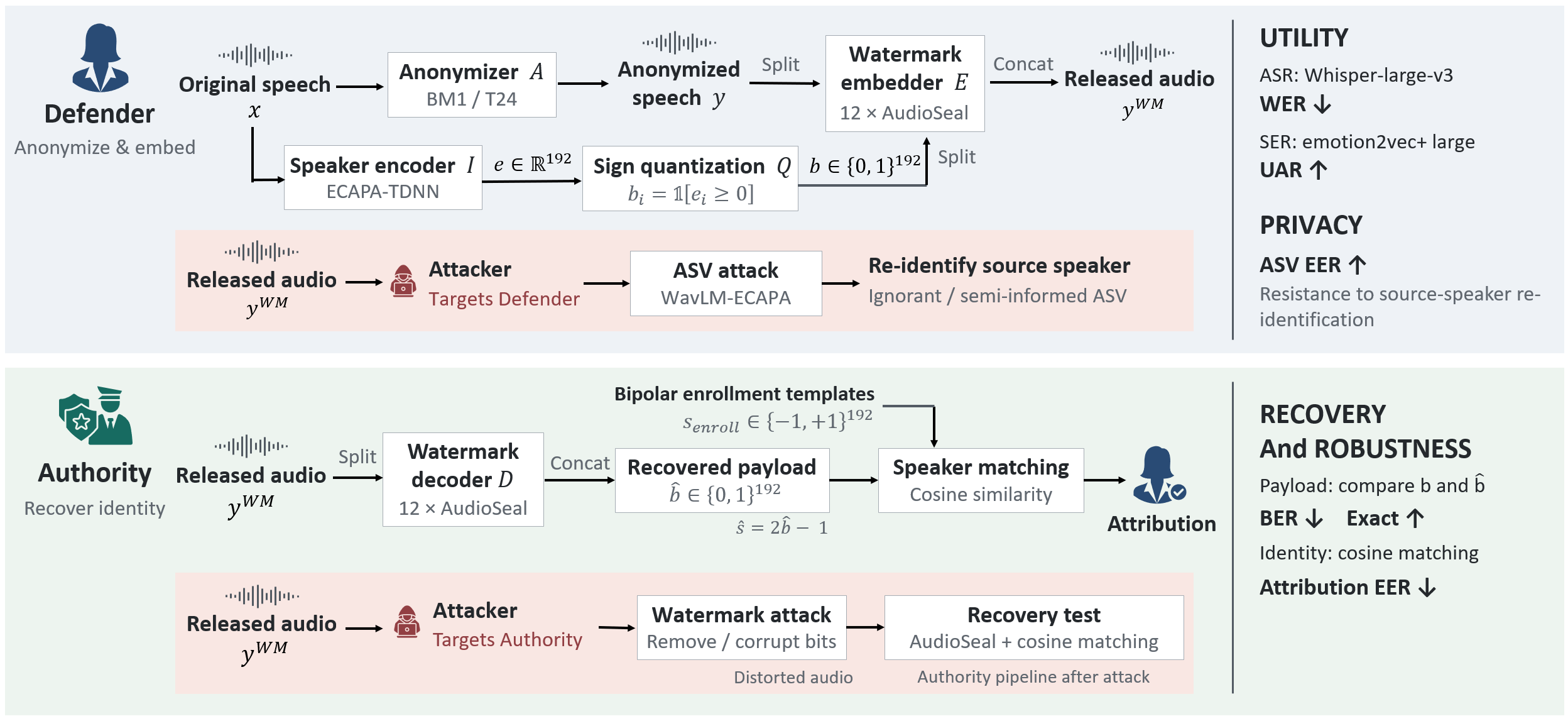}
    \caption{Overview of the proposed restorable anonymization pipeline.}
    \label{fig:pipeline}
\end{figure*}

\section{Related Work}

Our work lies at the intersection of multilingual voice anonymization and reversible speech privacy. The former investigates how to conceal speaker identity while preserving linguistic and paralinguistic information across languages, whereas the latter explores mechanisms for recovering original speech or speaker identity information from privacy-protected recordings. Together, these two research directions provide the foundation for our study of restorable multilingual voice anonymization.

\subsection{Multilingual Voice Anonymization}

Multilingual voice anonymization has been explored through both VC-based and ASR–TTS-based approaches. In VC-based anonymization, a key strategy is to replace language-specific ASR-based content extraction with self-supervised or language-independent representations. For example, \citet{miao2022language} used self-supervised speech representations to extract linguistic content without relying on language-specific ASR acoustic models. \citet{yao2025musa} further explored multilingual anonymization through serial disentanglement with semantic and speaker distillation. However, reducing language dependence in content extraction does not eliminate mismatches elsewhere in the pipeline: language and channel differences can still affect synthesis, motivating multilingual vocoder training and domain adaptation \cite{miao2022analyzing}.

ASR–TTS-based anonymization instead supports multiple languages through multilingual speech recognition and synthesis components. \citet{meyer2022speaker} investigated anonymization using phonetic intermediate representations, and subsequent work extended this approach to nine languages using multilingual ASR and TTS models \cite{meyer2024probing}. Beyond linguistic content, emotion-aware approaches incorporate emotional and prosodic information into resynthesis \cite{lee2024voice}. More recently, a reference-based approach combines multilingual transcription with zero-shot TTS, using emotion-matched reference utterances to guide both the voice and emotional characteristics of the anonymized output \cite{ferreira2026emotion}. 

Together, these studies establish two routes to multilingual anonymization: reducing language dependence in VC pipelines and using multilingual recognition and synthesis in ASR–TTS pipelines. Their architectural differences provide a useful setting for examining whether the same post-hoc identity recovery mechanism can operate across both approaches.

\subsection{Recoverable Speech Privacy}

Existing research on reversible speech privacy can be broadly categorized into voice restoration and speaker identity traceability. Voice restoration methods aim to reconstruct the original speech from its privacy-protected version. SpeechGuard \cite{zhang2025speechguard} employs reversible acoustic transformations and encryption of sensitive content, enabling fine-grained, permission-based recovery. Rano \cite{wang2025rano} instead utilizes conditional invertible neural networks to anonymize speech and restore the original waveform with a designated key, without requiring precise feature disentanglement. While these approaches support comprehensive recovery of protected speech, they rely on specifically designed reversible anonymization mechanisms, limiting their direct applicability to existing anonymization systems.

Speaker identity traceability offers an alternative by preserving identity information without necessarily reconstructing the original speech. VoxTracer \cite{ren2023speaking} integrates source-speaker embeddings into voice conversion through a VAE--Glow framework, enabling robust speaker tracing even after lossy audio compression. However, its application focuses on voice conversion rather than broader anonymization paradigms. More recently, \citet{ge2025proactive} embedded speaker representations as neural watermarks to detect identity manipulation by comparing the recovered representation with that of the received speech. Although effective under various signal transformations, its detection performance degrades under neural codec processing, and its objective differs from identifying the original speaker of anonymized speech.

While the former direction offers detailed speech restoration and the latter enables more compact identity tracing, neither has established a general post-hoc approach to speaker attribution across different multilingual anonymization systems. We address this gap by embedding a recoverable speaker representation into already anonymized speech, without reconstructing the original waveform or modifying the underlying anonymization models.

\section{Methodology}

We first formulate the recoverable voice anonymization task and its threat model, then describe our watermark-based identity attribution framework, including the anonymization backbones, binary speaker representation, and watermark embedding and recovery.

\subsection{Recoverable Voice Anonymization Task}
\label{sec:task}

As illustrated in Figure~\ref{fig:pipeline}, we extend the voice anonymization setting to include a \emph{defender}, a \emph{trusted authority}, and an \emph{attacker}. The defender releases anonymized utterances for downstream tasks such as ASR and SER, while the trusted authority may need to identify the original speakers for auditing. The attacker seeks either to infer speaker identity without authorization or to disrupt identity recovery.

The released recordings are referred to as \emph{trial} utterances. The defender first processes each utterance with an anonymization system to conceal its speaker identity, then embeds recoverable speaker information into the anonymized output. The trusted authority extracts this information and compares it with speaker templates derived from independent \emph{enrollment} recordings. Recovery therefore operates at the level of identity attribution rather than reconstruction of the original voice or waveform.

The task has four objectives:
(a)~to output an anonymized speech waveform;
(b)~to retain the underlying anonymization system's privacy and utility performance;
(c)~to enable speaker attribution through the recovery mechanism;
and (d)~to support different anonymization systems and languages without modifying the underlying anonymization models.

We consider two types of adversaries with different objectives. The \textit{privacy attacker} attempts to infer speaker identity from the released anonymized speech using automatic speaker verification (ASV), without explicitly decoding the watermark. We consider both a lazy-informed attacker, which is not adapted to the anonymization system, and a semi-informed attacker, which is trained on anonymized speech. The latter is evaluated under different watermarking conditions to examine whether watermark insertion introduces exploitable speaker information.

The \textit{watermark attacker} instead aims to disrupt authorized identity recovery by modifying the released watermarked speech. We consider common audio transformations, including filtering, additive noise, temporal modifications, and lossy compression, and assess whether the embedded payload remains recoverable and useful for speaker attribution. These transformations model potential watermark disruption rather than an adaptive, decoder-aware attack.

The trusted authority is assumed to have exclusive access to the watermark decoder and access to enrolled speaker representations. Exclusive decoder access represents a deployment-level trust assumption rather than an enforced security property of our implementation, which uses publicly available pretrained watermarking models. An attacker with access to the decoder and suitable reference recordings could therefore perform the same identity matching as the authority.

\subsection{Watermark-Based Identity Attribution}

We implement identity attribution by embedding a binary speaker representation into anonymized speech as an audio watermark. The upper part of Figure~\ref{fig:pipeline} illustrates the anonymization and watermark embedding pipeline. Given an input utterance $x$, an anonymization system $A$ first produces anonymized speech $y=A(x)$. In parallel, a pretrained speaker encoder $I$ extracts a speaker embedding $e$ from the original speech. A sign quantization function $Q$ converts $e$ into a fixed-length binary payload, with one bit per embedding dimension. This payload is embedded into $y$ using an audio watermarking model $E$, producing the final watermarked anonymized speech:
\begin{equation}
y^{\mathrm{WM}} = E\big(A(x), Q(I(x))\big).
\label{eq:overview_embedding}
\end{equation}

The lower part of Figure~\ref{fig:pipeline} illustrates identity recovery. A watermark decoder $D$ extracts a binary payload $\hat{b}=D(y^{\mathrm{WM}})$ from the released speech. We map the recovered bits to a bipolar representation $\hat{s}=2\hat{b}-\mathbf{1}$, converting 0 and 1 to $-1$ and $+1$, respectively, so that cosine similarity treats matching positive and negative signs symmetrically. Speaker matching is then performed using cosine similarity:
\begin{equation}
\mathrm{score}(y^{\mathrm{WM}}, s_{\mathrm{enroll}})
= \cos\big(2D(y^{\mathrm{WM}})-\mathbf{1}, s_{\mathrm{enroll}}\big),
\label{eq:overview_recovery}
\end{equation}
where $s_{\mathrm{enroll}}$ is the bipolar enrollment template, obtained by averaging a speaker's continuous enrollment embeddings, applying sign quantization, and mapping the resulting bits to $\{-1,+1\}$.

\subsubsection{Anonymizers}

We consider two multilingual voice anonymization backbones representing different technical paradigms: BM1, the official VPC 2026 multilingual track baseline \cite{miao2022language}, and a local reproduction of T24, an emotion-aware reference-based ASR-TTS system \cite{ferreira2026emotion}. 

BM1 follows a representation-based voice conversion architecture that separates linguistic content, prosody, and speaker identity. It extracts the fundamental frequency (F0) using YAAPT \cite{zahorian2008spectral} and content representations using a HuBERT-based soft content encoder \cite{hsu2021hubert}. An ECAPA-based speaker encoder \cite{desplanques2020ecapa} extracts speaker embeddings, which are replaced with pseudo-speaker x-vectors selected from a speaker pool to conceal the original identity. Finally, a HiFi-GAN vocoder \cite{kong2020hifi} synthesizes anonymized speech conditioned on the extracted content, F0, and pseudo-speaker representation. 

T24 adopts an emotion-aware ASR--TTS pipeline that regenerates speech from recognized text and an anonymizing voice reference. Whisper \cite{radford2023robust} first transcribes the input speech, while emotion2vec \cite{ma2024emotion2vec} extracts an emotion embedding. This embedding is compared with those of candidate utterances in a language-specific voice bank using cosine similarity. A reference is randomly selected from the 50 most emotionally similar candidates to preserve emotional characteristics while introducing speaker diversity. VoxCPM2 \cite{zhou2026voxcpm2} then synthesizes the anonymized speech conditioned on the recognized text and the selected reference speech.

\begin{figure}[t]
    \centering
    \includegraphics[width=0.95\linewidth]{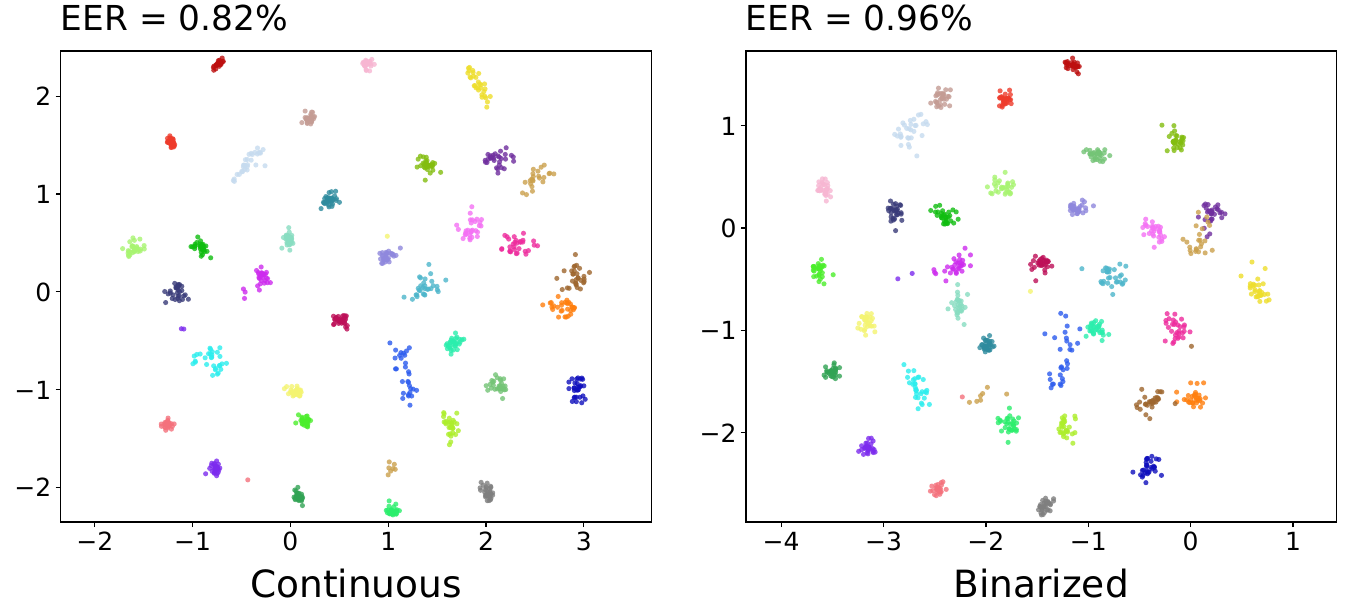}
    \caption{t-SNE of continuous and sign-quantized speaker embeddings. EER values are unweighted four-language means on the full test set.}
    \label{fig:tsne}
\end{figure}

\begin{table}[t]
    \centering
    \small
    \begin{tabular}{lll}
        \toprule
        Language & Continuous $\downarrow$ & Binary $\downarrow$ \\
        \midrule
        EN  & 2.761 & 2.919 \\
        FR  & 0.093 & 0.145 \\
        DE  & 0.244 & 0.495 \\
        ES  & 0.197 & 0.270 \\
        \rowcolor{gray!20}
        AVG & 0.824 & 0.957 \\
        \bottomrule
    \end{tabular}
    \caption{Speaker attribution EER (\%) using continuous and sign-quantized embeddings across four languages.}
    \label{tab:eer_comparison}
\end{table}

\subsubsection{Binary Speaker Representation}

Prior studies have explored binary speaker embeddings for efficient speaker verification and privacy-preserving biometric matching \cite{li2016binary,mtibaa2021privacy}. We adopt zero-threshold sign quantization to convert continuous speaker embeddings into binary payloads suitable for audio watermarking, retaining their sign patterns for subsequent speaker matching.

Given a 192-dimensional speaker embedding $e$ extracted by a pretrained ECAPA-TDNN encoder \cite{desplanques2020ecapa}, we define
\begin{equation}
b_i = \mathbf{1}[e_i \ge 0], \qquad i=1,\ldots,192,
\end{equation}
yielding a payload $b\in\{0,1\}^{192}$. This quantization retains the sign of each embedding dimension while discarding its magnitude. For speaker matching, we map the bits to the bipolar representation $s=2b-\mathbf{1}\in\{-1,+1\}^{192}$.

We evaluate the effect of discarding embedding magnitudes by comparing continuous and binary representations in the experiments.

\begin{table*}[t]
\centering
\caption{Privacy and utility evaluation on the VPC 2026 Track 2 test sets.}
\small
\label{tab:privacy_utility}
\renewcommand{\arraystretch}{1.15}
\resizebox{\textwidth}{!}{
\begin{tabular}{
    l
    *{4}{c}>{\columncolor[gray]{0.94}}c
    *{4}{c}>{\columncolor[gray]{0.94}}c
    *{4}{c}>{\columncolor[gray]{0.94}}c
}
\toprule
\multirow{2}{*}{System}
& \multicolumn{5}{c}{SER UAR $\uparrow$}
& \multicolumn{5}{c}{ASR WER $\downarrow$}
& \multicolumn{5}{c}{Lazy-informed ASV EER $\uparrow$} \\
\cmidrule(lr){2-6}
\cmidrule(lr){7-11}
\cmidrule(lr){12-16}
& HAP & NEU & ANG & SAD & AVG
& EN & FR & DE & ES & AVG
& EN & FR & DE & ES & AVG \\
\midrule
Original
& 68.077 & 90.638 & 83.396 & 73.504 & 78.904
& 5.469 & 5.141 & 5.379 & 3.761 & 4.938
& 4.205 & 6.475 & 3.276 & 3.027 & 4.246 \\
\midrule
BM1
& 38.846 & 70.638 & 11.321 & 44.017 & 41.206
& 8.125 & 9.688 & 9.349 & 6.055 & 8.304
& 4.306 & 9.302 & 8.961 & 5.949 & 7.130 \\
BM1 + WM
& 38.462 & 72.340 & 10.943 & 45.299 & 41.761
& 8.157 & 9.699 & 9.349 & 5.954 & 8.290
& 4.454 & 9.542 & 9.341 & 6.148 & 7.371 \\
\midrule
T24
& 36.154 & 70.213 & 62.642 & 56.838 & 56.461
& 9.104 & 9.515 & 7.216 & 8.058 & 8.473
& 38.688 & 47.656 & 44.998 & 48.334 & 44.919 \\
T24 + WM
& 36.538 & 69.787 & 63.396 & 56.838 & 56.640
& 8.504 & 9.378 & 7.042 & 7.920 & 8.211
& 38.757 & 47.901 & 45.002 & 48.292 & 44.988 \\
\bottomrule
\end{tabular}
}
\end{table*}

\begin{table}[t]
    \centering
    \caption{Semi-informed ASV EER (\%) on the T24 test sets.
    $\checkmark$ and $\times$ indicate whether watermarked anonymized speech is used for ASV training.}
    \label{tab:t24_semi_asv}
    \small
    \renewcommand{\arraystretch}{1.15}
    \setlength{\tabcolsep}{4pt}
    \begin{tabular}{llccccc}
        \toprule
        \multirow{2}{*}{System}
        & \multirow{2}{*}{WM}
        & \multicolumn{5}{c}{EER (\%) $\uparrow$} \\
        \cmidrule(lr){3-7}
        & & EN & FR & DE & ES
        & \cellcolor[gray]{0.94} AVG \\
        \midrule
        \multirow{2}{*}{T24}
        & $\times$ & 31.064 & 45.131 & 39.909 & 43.480
        & \cellcolor[gray]{0.94} 39.896 \\
        & $\checkmark$ & 31.001 & 44.791 & 41.094 & 43.332
        & \cellcolor[gray]{0.94} 40.055 \\
        \midrule
        \multirow{2}{*}{T24 + WM}
        & $\times$ & 31.221 & 45.031 & 39.700 & 44.274
        & \cellcolor[gray]{0.94} 40.057 \\
        & $\checkmark$ & 31.001 & 44.304 & 39.936 & 42.985
        & \cellcolor[gray]{0.94} 39.557 \\
        \bottomrule
    \end{tabular}
\end{table}

\begin{table}[t]
    \centering
    \caption{Robustness of WM12 under audio distortions.}
    \label{tab:robustness}
    \setlength{\tabcolsep}{3pt}
    \renewcommand{\arraystretch}{1.1}
    \small
    \begin{tabular}{lrrrrrr}
        \toprule
        \multirow{2}{*}{Condition}
        & \multicolumn{2}{c}{BER (\%) $\downarrow$}
        & \multicolumn{2}{c}{Exact (\%) $\uparrow$}
        & \multicolumn{2}{c}{Attr EER (\%) $\downarrow$} \\
        \cmidrule(lr){2-3}
        \cmidrule(lr){4-5}
        \cmidrule(lr){6-7}
        & BM1 & T24 & BM1 & T24 & BM1 & T24 \\
        \midrule
        None
        & 0.005 & 0.008 & 99.362 & 98.989
        & 0.963 & 0.956 \\
        \midrule
        Band-pass
        & 0.004 & 0.011 & 99.573 & 98.581
        & 0.964 & 0.957 \\
        High-pass
        & 0.004 & 0.011 & 99.568 & 98.611
        & 0.964 & 0.957 \\
        Low-pass
        & 0.005 & 0.008 & 99.352 & 99.018
        & 0.963 & 0.956 \\
        \midrule
        Pink noise
        & 0.005 & 0.009 & 99.372 & 98.920
        & 0.963 & 0.956 \\
        White noise
        & 0.312 & 0.446 & 69.298 & 62.709
        & 0.979 & 0.971 \\
        \midrule
        Speed
        & 50.274 & 50.577 & 0.000 & 0.000
        & 48.952 & 49.363 \\
        Resample
        & 0.005 & 0.008 & 99.362 & 98.989
        & 0.963 & 0.956 \\
        \midrule
        Boost
        & 0.004 & 0.009 & 99.519 & 98.900
        & 0.963 & 0.957 \\
        Duck
        & 0.015 & 0.008 & 97.614 & 98.984
        & 0.964 & 0.956 \\
        Echo
        & 0.588 & 1.338 & 44.523 & 52.978
        & 0.977 & 1.185 \\
        Smooth
        & 1.356 & 0.169 & 14.116 & 77.454
        & 0.965 & 0.957 \\
        \midrule
        AAC
        & 0.008 & 0.019 & 98.748 & 97.388
        & 0.963 & 0.957 \\
        MP3
        & 0.052 & 0.091 & 91.413 & 89.095
        & 0.970 & 0.954 \\
        \bottomrule
    \end{tabular}
\end{table}

\subsubsection{Watermark Embedding and Recovery}

We embed the binary identity payload using AudioSeal \citep{roman2024proactive}. AudioSeal employs an EnCodec-inspired convolutional encoder--decoder generator that produces a message-conditioned additive watermark, together with a convolutional detector for watermark localization and message recovery. The two components are jointly trained with perceptual and localization losses to improve watermark imperceptibility and detectability. 

Given anonymized speech $y$ sampled at 16 kHz, we partition the waveform into 12 contiguous segments, each carrying a 16-bit chunk of the 192-bit payload. For a waveform of $N=12q+r$ samples ($0\leq r<12$), the first $r$ segments contain $q+1$ samples each, while the remaining segments contain $q$ samples. For batched AudioSeal inference, shorter segments are temporarily zero-padded to equal length. A frozen AudioSeal generator embeds the corresponding message chunk into each segment, after which the outputs are trimmed to their original segment lengths and concatenated to form $y^{\mathrm{WM}}$, preserving the original waveform length.

For recovery, the decoder applies the same 12-way partition to $y^{\mathrm{WM}}$ and extracts one 16-bit message from each segment. The recovered chunks are concatenated to obtain $\hat b$.

\begin{figure*}[t]
    \centering
    \includegraphics[width=0.85\linewidth]{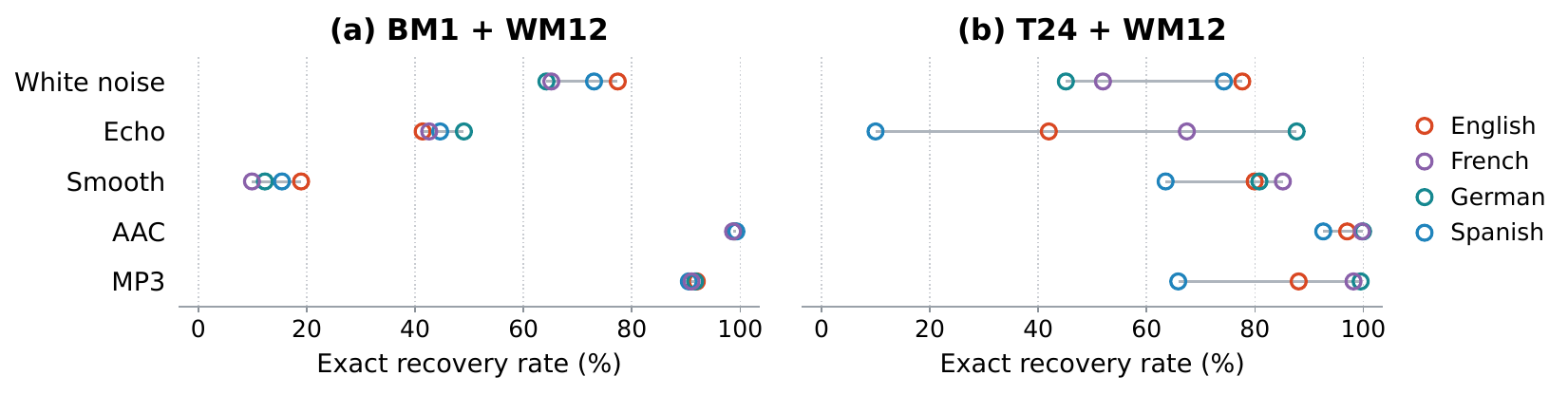}
    \caption{Language-dependent exact watermark recovery
    under representative audio distortions.}
    \label{fig:language_robustness}
\end{figure*}

\section{Experiments}

\subsection{Experimental Setup}

\paragraph{Data}

We follow the multilingual anonymization protocol of VPC 2026, covering English, French, German, and Spanish \cite{miao:hal-05561895}. ASV privacy and ASR utility are evaluated on test sets drawn from Multilingual LibriSpeech (MLS) \cite{Pratap2020MLSAL}, while emotional preservation is assessed on a separate test set containing 994 utterances spanning the same four languages and four emotion classes: happiness, neutrality, anger, and sadness. Speaker attribution is evaluated on the MLS test sets using the corresponding enrollment recordings, whereas watermark recovery metrics are computed over the combined MLS development and test sets.

\paragraph{Anonymizers}
We use the official VPC 2026 implementation of BM1\footnote{\url{https://github.com/Voice-Privacy-Challenge/Voice-Privacy-Challenge-2026/blob/main/anonymization/pipelines/ssl}} \cite{miao2022language} and locally reproduce T24 following \citet{ferreira2026emotion}. The component implementations are available for Whisper-large-v3\footnote{\url{https://huggingface.co/openai/whisper-large-v3}} \cite{radford2023robust}, emotion2vec+ large\footnote{\url{https://huggingface.co/emotion2vec/emotion2vec_plus_large}} \cite{ma2024emotion2vec}, and VoxCPM2\footnote{\url{https://huggingface.co/openbmb/VoxCPM2}} \cite{zhou2026voxcpm2}. For the emotion reference bank we constructed, the English bank combines MSP-Podcast\footnote{\url{https://www.lab-msp.com/MSP/MSP-Podcast.html}} \cite{busso2026msp}, CAMEO\footnote{\url{https://huggingface.co/datasets/amu-cai/CAMEO}} \cite{christop2026cameo}, and a subset of MEAD\footnote{\url{https://huggingface.co/datasets/NoahMartinezXiang/2020_MEAD}} \cite{wang2020mead}, while the French, German, and Spanish banks use their respective CAMEO language subsets.

For identity representation, we use a pretrained 192-dimensional ECAPA-TDNN encoder \cite{desplanques2020ecapa}. AudioSeal's pretrained generator and detector \citep{roman2024proactive} are used for watermark embedding and recovery without additional fine-tuning.

\paragraph{Evaluation Metrics}
We follow the VPC 2026 evaluation protocols to compare each anonymization backbone before and after watermark insertion. Privacy is evaluated using a WavLM-ECAPA ASV system \cite{chen2022wavlm} and measured by ASV equal error rate (EER), where higher values indicate stronger privacy against the evaluated attackers. We evaluate lazy-informed ASV attackers for both backbones and additionally evaluate semi-informed attackers trained on T24 outputs with or without watermarking. Linguistic intelligibility is evaluated using Whisper-large-v3 and measured by word error rate (WER; lower is better). Emotional preservation is evaluated using emotion2vec+ large and measured by unweighted average recall (UAR; higher is better).

We additionally assess watermark recovery and speaker attribution under clean conditions and audio distortions. Watermark recovery is measured by bit error rate (BER), the proportion of incorrectly recovered bits, and exact payload recovery rate (Exact), the percentage of utterances with all 192 bits correctly recovered. Lower BER and higher Exact indicate better recovery. Speaker attribution is measured by attribution EER based on cosine similarity between recovered speaker representations and enrollment templates. Unlike ASV EER for privacy evaluation, lower attribution EER indicates better performance.

\subsection{Experimental Results}

\paragraph{Binary Representation Analysis}

We first examine whether sign quantization retains useful speaker-discriminative information. Figure~\ref{fig:tsne} presents speaker-colored t-SNE projections of continuous and binary speaker representations, using the same 1,050 original test utterances sampled from 35 speakers across four languages. Both representations exhibit visible speaker-dependent grouping, suggesting that the sign transformation retains part of the original embedding structure.

Table ~\ref{tab:eer_comparison} quantitatively evaluates the impact of sign quantization on speaker attribution across four languages. Compared with continuous embeddings, binary representations yield slightly higher EERs in all languages, with the average increasing from 0.8238\% to 0.9573\%. Despite discarding embedding magnitudes, the binary representations retain strong speaker-discriminative capability, supporting their use as compact 192-bit identity payloads.

\paragraph{Impact on Anonymization Privacy and Utility}

Table~\ref{tab:privacy_utility} shows that watermarking largely preserves each backbone's measured privacy-utility trade-off. ASV EER measures speaker verification difficulty, with higher values indicating stronger privacy. ASR WER measures speech recognition error, with lower values indicating better linguistic preservation. SER UAR measures the average recall across four emotion classes (HAP, NEU, ANG, and SAD), with higher values indicating better emotional preservation. The Original row represents unprocessed speech, while BM1 and T24 denote their respective anonymized outputs; +WM indicates post-hoc watermarking. ASV evaluation follows the VPC 2026 lazy-informed protocol, using OO for original speech and AA for anonymized speech. All values are percentages. AVG denotes the unweighted mean across four languages for ASV EER and ASR WER, and the unweighted mean across four emotion classes for SER UAR.

Overall, post-hoc watermarking has a negligible impact on both ASV privacy and ASR utility across the two anonymization backbones, suggesting that the watermark largely preserves their original privacy-utility trade-offs. The two systems exhibit different emotional preservation capabilities: T24 achieves substantially better SER performance than BM1, particularly for anger recognition. Nevertheless, both systems show a considerable reduction in emotional utility compared with original speech, indicating that this degradation primarily originates from the anonymization process rather than watermark insertion.

In Table~\ref{tab:t24_semi_asv}, the System column specifies the evaluation audio condition, while the WM column indicates whether watermarked T24 speech is used to train the semi-informed ASV attacker. AVG denotes the unweighted mean across four languages. The results show that watermark insertion has little impact on ASV privacy, with average EERs remaining between 39.557\% and 40.057\% across training and evaluation conditions.

\begin{figure*}[t]
    \centering
    \includegraphics[width=1\linewidth]{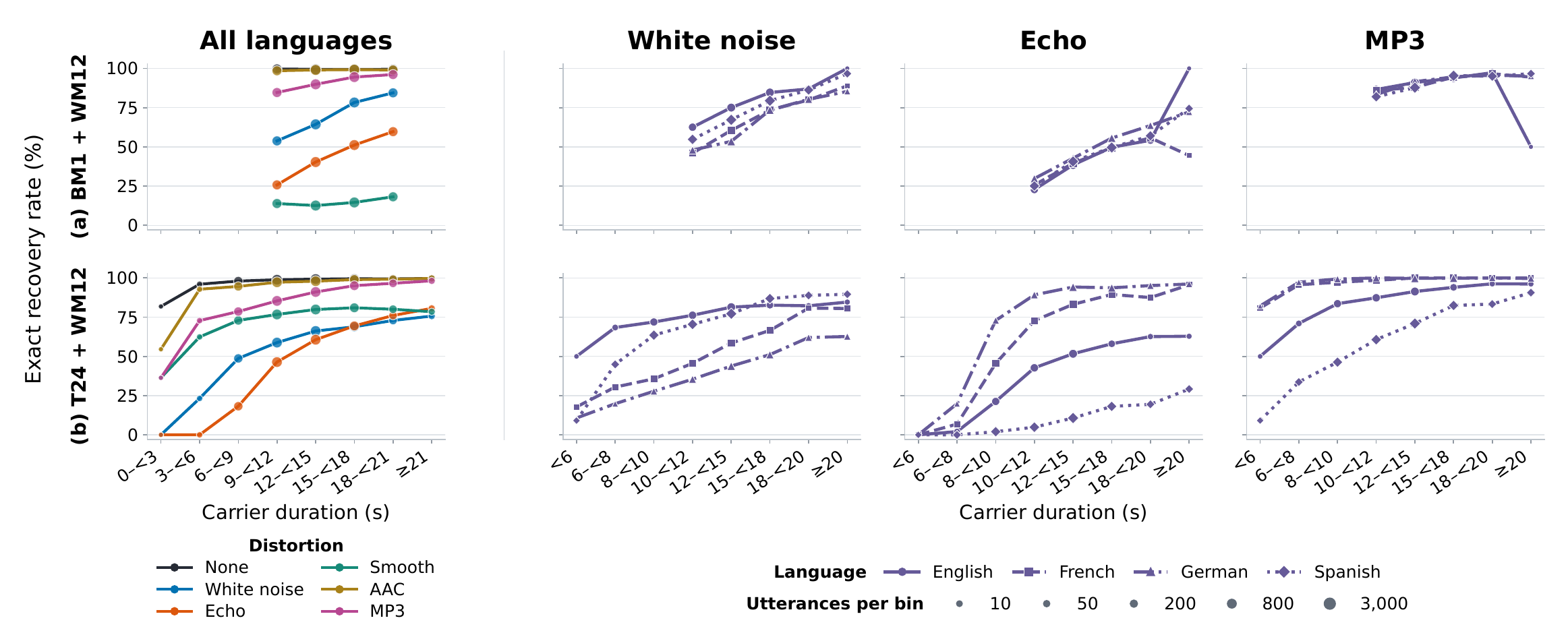}
    \caption{Duration-dependent exact watermark recovery
on representative audio distortions. The first column shows results across all languages, while the remaining columns compare over four languages. Results are computed on the VPC 2026 test sets.}
    \label{fig:duration_robustness}
\end{figure*}

\paragraph{Recovery and Robustness Analysis}

We apply each distortion to the watermarked trial waveform before decoding, leaving original enrollment unchanged. The fixed settings are band-pass filtering at 300--8000\,Hz, high-pass at 500\,Hz, low-pass at 5000\,Hz, pink and white noise with standard deviations 0.01 and 0.001, speed scaling by 1.25, resampling from 16 to 32 and back to 16\,kHz, amplitude boost/duck by 20\%, echo with 0.5\,s delay and 0.5 gain, a six-sample smoothing window, AAC at 64\,kbps, and MP3 at 32\,kbps. Noise is seeded per utterance, and neither synchronization recovery nor error correction is used.

Table~\ref{tab:robustness} evaluates payload recovery and speaker attribution under different audio distortions. BER measures the proportion of incorrectly recovered bits, while Exact denotes the percentage of utterances for which all 192 bits are correctly recovered. Attr EER measures speaker attribution performance by computing EER from cosine similarity scores between recovered test representations and enrollment templates, with lower values indicating better attribution. BER and Exact are pooled over development and test sets, whereas Attr EER is the unweighted mean across four test languages.

Without distortion, Exact reaches 99.362\% for BM1 and 98.989\% for T24, with attribution EERs of 0.963\% and 0.9563\%, respectively. As a reference, directly comparing sign-quantized representations extracted from original trial recordings against the same enrollment templates, without watermark embedding or recovery, gives 0.957\% mean EER. The close attribution results indicate that watermark embedding and recovery introduce little degradation under clean conditions, although they do not establish lossless binarization relative to continuous embeddings.

Most distortions have little effect on attribution despite lower Exact. White noise reduces T24 Exact to 62.7093\%, yet attribution EER remains 0.9709\%. Smoothing reduces BM1 Exact to 14.1160\%, while attribution EER remains 0.9652\%. Thus, exact payload reconstruction is stricter than speaker verification: partially corrupted representations can retain discriminative information. Speed scaling is the clear failure case: BER approaches 50\%, Exact falls to zero, and attribution EER reaches 48.9516\% for BM1 and 49.3629\% for T24. The current decoder, without speed compensation or synchronization search, is vulnerable to temporal rescaling.

\noindent
\textbf{Language-Dependent Robustness.} Figure ~\ref{fig:language_robustness} further examines exact watermark recovery across the four languages under representative distortions on the VPC 2026 test sets, where each marker indicates the percentage of utterances with all 192 payload bits correctly recovered for a given language and gray lines indicate the range across languages for each distortion. Under white noise, French and German exhibit lower recovery rates than English and Spanish for both anonymizers. More pronounced language-dependent differences emerge with T24: under echo, exact recovery ranges from 9.97\% in Spanish to 87.70\% in German. A similar pattern appears under lossy compression, where Spanish achieves 92.61\% and 65.84\% under AAC and MP3, respectively, compared with 99.97\% and 99.51\% for German. In contrast, BM1 exhibits smaller cross-language variations under these compression conditions. Smoothing reveals a different pattern, substantially reducing recovery for BM1 (9.88--18.95\%) while T24 remains relatively robust (63.51--85.15\%) across all languages. These findings suggest that watermark robustness depends jointly on language and anonymization backbone, potentially reflecting differences in the acoustic characteristics of their generated speech.

\noindent
\textbf{Duration-Dependent Robustness.} We further investigate how carrier duration affects watermark recovery under different audio distortions. As shown in Figure~\ref{fig:duration_robustness}, exact recovery generally improves with increasing duration, particularly under white noise, echo, and MP3 compression. This trend is consistent with our fixed 12-segment embedding strategy, where longer recordings provide more audio samples for each 16-bit watermark chunk. The two anonymization backbones exhibit different sensitivities: T24 shows substantial recovery improvements with longer carriers, especially under echo, whereas BM1 maintains low recovery rates under smoothing even for longer recordings. In contrast, recovery under clean conditions and AAC compression remains relatively high across most duration groups. These observations suggest that carrier duration is an important factor in watermark robustness, but its effect depends on both the distortion type and the anonymization backbone. 

The language-specific panels in Figure~\ref{fig:duration_robustness} further examine multilingual robustness across comparable carrier-duration groups. While recovery generally improves with duration, substantial cross-language differences persist for T24. Under echo, German maintains considerably higher recovery than Spanish across comparable duration groups, with the gap remaining pronounced even for longer utterances. Under MP3 compression, however, Spanish shows substantial improvement with increasing duration, gradually approaching the other languages. In contrast, BM1 exhibits relatively small cross-language variations under white noise and MP3 compression. These observations suggest that carrier duration alone does not fully account for the language-dependent recovery differences, which also vary with distortion type and anonymization backbone.

\section{Conclusion}

In this work, we presented a post-hoc framework for restorable multilingual voice anonymization that enables speaker attribution without reconstructing the original waveform or modifying the underlying anonymization systems. By combining sign-quantized speaker embeddings with segment-wise audio watermarking, our approach embeds recoverable identity information into anonymized speech. Experiments across four languages and two distinct anonymization backbones demonstrate near-99\% exact payload recovery and approximately 0.96\% attribution EER under clean conditions, while largely preserving the original privacy--utility trade-offs. Moreover, speaker attribution remains effective under several audio distortions despite imperfect payload recovery. Further analyses reveal that watermark robustness varies across languages, carrier durations, and anonymization backbones.

Nevertheless, the fixed-segment embedding strategy remains vulnerable to temporal rescaling, and the use of a publicly available watermark detector does not enforce access control over embedded identities. Future work will investigate synchronization-aware watermark recovery, secure identity access mechanisms, and broader evaluation across anonymization systems and languages.

\bibliography{aaai2027}


\end{document}